\documentclass[letterpaper,twocolumn,english,aps,prl,superscriptaddress]{revtex4-1}
\usepackage[latin9]{inputenc}
\usepackage{amsmath}
\usepackage{amssymb}
\usepackage{stackrel}
\usepackage{graphicx}

\makeatletter

\pdfpageheight\paperheight
\pdfpagewidth\paperwidth

\usepackage{sidecap}
\usepackage{xcolor}
\definecolor{darkblue}{rgb}{0.1,0.2,0.6} 
\definecolor{lightblue}{rgb}{0.1,0.1,1.0}
\definecolor{darkred}{rgb}{0.8,0.1,0.2}
\usepackage[colorlinks,citecolor=lightblue,linkcolor=darkblue,urlcolor=lightblue] {hyperref}

\makeatother

\usepackage{babel}
\begin{document}
\title{Stark-Many body localization in interacting infinite dimensional systems }
\author{Hristiana Atanasova}
\affiliation{School of Chemistry, Tel Aviv University, Tel Aviv 6997801, Israel}
\author{André Erpenbeck}
\affiliation{Department of Physics, University of Michigan, Ann Arbor, Michigan
48109, USA}
\author{Emanuel Gull}
\affiliation{Department of Physics, University of Michigan, Ann Arbor, Michigan
48109, USA}
\author{Yevgeny Bar Lev}
\affiliation{Department of Physics, Ben-Gurion University of the Negev, Beer-Sheva
84105, Israel}
\author{Guy Cohen}
\email{gcohen@tau.ac.il}

\affiliation{School of Chemistry, Tel Aviv University, Tel Aviv 6997801, Israel}
\affiliation{The Raymond and Beverley Sackler Center for Computational Molecular
and Materials Science, Tel Aviv University, Tel Aviv 6997801, Israel}
\date{\today}
\begin{abstract}
We study bulk particle transport in a Fermi\textendash Hubbard model
on an infinite-dimensional Bethe lattice, driven by a constant electric
field. Previous numerical studies showed that one dimensional analogs
of this system exhibit a breakdown of diffusion due to Stark many-body
localization (Stark-MBL) at least up to time which scales exponentially
with the system size. Here, we consider systems initially in a spin
density wave state using a combination of numerically exact and approximate
techniques. We show that for sufficiently weak electric fields, the
wave's momentum component decays exponentially with time in a way
consistent with normal diffusion. By studying different wavelengths,
we extract the dynamical exponent and the generalized diffusion coefficient
at each field strength. Interestingly, we find a non-monotonic dependence
of the dynamical exponent on the electric field. As the field increases
towards a critical value proportional to the Hubbard interaction strength,
transport slows down, becoming sub-diffusive. At large interaction
strengths, however, transport speeds up again with increasing field,
exhibiting super-diffusive characteristics when the electric field
is comparable to the interaction strength. Eventually, at the large
field limit, localization occurs and the current through the system
is suppressed.
\end{abstract}
\maketitle

Isolated, interacting quantum systems with many degrees of freedom
generically approach thermal equilibrium at least for local observables.
One of the few exceptions to this is the breaking of ergodicity by
sufficiently strong disorder, which leads to many-body localization
(MBL) \citep{Basko2006,Gornyi2005,Nandkishore2014,Abanin2018,Alet2018}.
Advances in ultracold atomic experiments have enabled observation
of the MBL phase, as well as the study of its dynamical properties
and its response to external probes \citep{Schreiber2015,Choi2016}.
Much of this work is driven by technological promise: MBL suppresses
heating of periodically driven system \citep{Abanin2014,Lazarides2014,Ponte2014a},
and may therefore be useful in the design of quantum information storage
devices.

MBL manifests the stability of the \emph{noninteracting} Anderson
insulator \citep{Anderson1958b} at sufficiently small interactions.
Localization in noninteracting systems is, however, \emph{not} limited
to disordered systems. For example, single-particle states can be
localized by a spatially uniform ac electric field, an effect known
as dynamic localization \citep{Dunlap1986,Dunlap1988}; however, typically
such localization mechanisms are unstable to the addition of interactions
\citep{Luitz2017b}. Conversely, numerical studies \citep{Schulz2019,vanNieuwenburg2019}
and cold atom experiments \citep{Guardado-Sanchez2020,Scherg2021,morong2021:ObservationStarkManybody}
have shown that\textemdash in the presence of a static and spatially
uniform dc electric field\textemdash localization can exhibit a robustness
to interactions. This phenomenon has been dubbed Stark-MBL. Nevertheless,
recent studies have shown that localization might only persist up
to a finite timescale controlled by the size of the system~\citep{zisling2022:TransportStarkManybody,kloss2023:AbsenceLocalizationInteracting,gunawardana2022:DynamicalLbitsStark}.
It is therefore a largely open question whether Stark-MBL persists
in the thermodynamic limit (TDL).

Work on MBL is mostly focused on low-dimensional systems, partially
due to the availability of powerful numerical techniques for one-dimensional
systems, but also because Anderson localization occurs only within
states bounded by a mobility edge that shrinks with increasing dimension.
Noninteracting Stark localization, on the other hand, can exist in
parallel to the direction of the electric field at any dimension \citep{Wannier1960}.
Due to methodological constraints, theoretical studies of stability
with respect to interaction in dimensions higher than one have been
limited to perturbative approaches \citep{Zhang2020}. Experimental
study of a two-dimensional interacting Stark-MBL system suggests that
the system is delocalized and sub-diffusive \citep{Guardado-Sanchez2020,Zhang2020}.

The limit of infinite dimensions is accessible by way of the dynamical
mean field theory (DMFT) \citep{metzner_correlated_1989,georges_hubbard_1992,Georges_Dynamical_1996}.
This has enabled studies of the formation of long-lived, quasi-stationary
currents in the presence of uniform electric fields \citep{eckstein_dielectric_2010},
which in closed systems are eventually expected to decay due to heating
at very long timescales \citep{mierzejewski_nonlinear_2010}. The
effect of weak interactions on the Bloch oscillations characterizing
noninteracting Wannier\textendash Stark physics has also been studied
by such means \citep{eckstein_damping_2011}.

In this Letter, we consider the nonequilibrium dynamics \emph{directly}
at the thermodynamic limit of an infinite-dimensional Hubbard model
in the presence of a constant electric field. We show that transport
is inconsistent with generalized diffusion for sufficiently strong
electric fields, indicating a transition to a localized phase. Moreover,
we find a non-monotonic dependence of the dynamical transport exponent
on field strength: the system goes from diffusive behavior to subdiffusion,
then exhibits a superdiffusive resonant phase before becoming fully
localized at high field. We show that the superdiffusive behavior
coincides with increases in the quasistatic currents flowing through
the system, and argue that it is therefore a transient nonequilbrium
effect. Our results suggest that nonequilibrium physics at intermediate
timescales plays an important role in many-body localization, particularly
when large, high-dimensional systems are studied.

\paragraph{Model.\textemdash{}}

\begin{figure}
\includegraphics{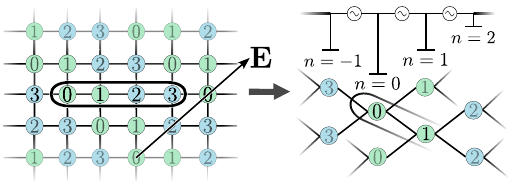}\caption{Model illustration. Circles represent Hubbard sites, with different
colors representing different spin directions in the initial state.
Lines denote hopping terms. The curved outline encompasses a unit
cell, and the arrow points along the electric field. (left) 2D model
with an initial condition having a periodicity of $l=4$. (right)
corresponding $l=4$ Bethe lattice with same proportions of neighboring
states with each initial condition. Layers with identical electrical
potential are arranged in vertical lines.\label{fig:model}}
\end{figure}

We investigate a particle\textendash hole-symmetric Hubbard model
describing fermions on a Bethe lattice with coordination number $Z$,
at the limit $Z\rightarrow\infty$:

\begin{equation}
\begin{aligned}\hat{H} & =-J\sum_{\left\langle ij\right\rangle \sigma}\hat{d}_{i\sigma}^{\dagger}\hat{d}_{j\sigma}+U\sum_{i}\left(\hat{n}_{i\uparrow}-\frac{1}{2}\right)\left(\hat{n}_{i\downarrow}-\frac{1}{2}\right).\end{aligned}
\label{eq:Hubbard_hamiltonian}
\end{equation}
Here, $\hat{d}_{i\sigma}^{(\dagger)}$ are fermionic annihilation(creation)
operators associated with lattice site $i$ and spin $\sigma$, and
$\hat{n}_{i\sigma}\equiv\hat{d}_{i\sigma}^{\dagger}\hat{d}_{i\sigma}$.
The fermions interact through a local Coulomb repulsion of strength
$U$, and can hop between neighboring lattice sites $\langle ij\rangle$
with hopping amplitude $J$. The hopping is defined such that it is
rescaled with respect to a bare hopping parameter $t_{0}$: $J\equiv\frac{t_{0}}{\sqrt{Z}}$.
We set $t_{0}\equiv1$ to be our unit of energy, and $\hbar\equiv1$.
As we explain below, our objective is to study generalized diffusion
in this model by simulating its relaxation from periodic, nonequilibrium
initial conditions characterized by discrete wavelengths; in the presence
of constant, uniform electric fields applied in parallel to the waves.

We parameterize the Bethe lattices to reflect the structure of a 2D
cubic lattice with the field applied along a main diagonal. The spatially
homogeneous electric field enters the Hamiltonian by means of the
Peierls substitution, using a pure time-dependent vector potential
$\mathbf{E}\left(t\right)=-\partial_{t}\mathbf{A}\left(t\right)$.
This introduces a potential difference between different diagonal
layers that can be expressed as a phase in the hopping amplitudes
$t_{0}$. As an initial state, except where stated otherwise, we chose
the spin density wave density matrix
\begin{equation}
\hat{\rho}_{\mathbf{k}}\left(0\right)=\otimes_{n}\left(\sin^{2}\left(\mathbf{k}n\right)\left|\uparrow\right\rangle \left\langle \uparrow\right|+\cos^{2}\left(\mathbf{k}n\right)\left|\downarrow\right\rangle \left\langle \downarrow\right|\right)_{n},\label{eq:initial_state}
\end{equation}
where $\mathbf{k}$ is a wave-vector in the first Brillouin zone.
We only consider $\mathbf{k}$'s that are parallel to the electric
field (see left panel of Fig.~\ref{fig:model}), and therefore are
fully defined by their wave number $k=\left|\mathbf{k}\right|$.

The 2D model could now be \emph{approximately} solved within nonequilibrium
DMFT \citep{Freericks_Nonequilibrium_2006,Aoki_Nonequilibrium_2014,Turkowski_Nonequilibrium_2021},
with a variety of generalizations available to provide systematic
corrections \citep{maier_quantum_2005}. Instead, we will construct
an analogous infinite-dimensional model for which DMFT is \emph{exact},
and which captures many of the relevant physical properties of the
2D system. Since the initial condition is periodic and the Hamiltonian,
Eq.~(\ref{eq:Hubbard_hamiltonian}), is translationally invariant
in the time-dependent gauge, it is sufficient to solve for the dynamics
of $l=\pi/k$ unique sites, which we will call a unit cell. Fig.~\ref{fig:model}
demonstrates the construction of the electric field for the Bethe
lattice with infinite coordination number and same properties as the
finite-dimensional model; the $Z=4$ version is illustrated in the
right panel of Fig.~\ref{fig:model}. While one could also use, e.g.,
a hypercubic lattice with the field on the diagonal \citep{Georges_Dynamical_1996},
this has a minor effect on the physics but somewhat complicates the
DMFT self-consistency condition.

\paragraph{Numerical solution.\textemdash{}}

The DMFT maps the extended interacting lattice model onto a set of
$l$ effective impurity models\textemdash one for each unique site\textemdash that
are coupled only by a self-consistency condition \citep{Georges_Dynamical_1996}.
We solve the auxiliary models by three different methods of increasing
complexity and precision, all of which are based on perturbative expansions
in the impurity\textendash bath hybridization. We rely mostly on the
non-crossing approximation (NCA) and one-crossing approximation (OCA),
which represent the lowest and next-to-lowest order self-consistent,
conserving truncations in the hybridization expansion \citep{Bickers_Review_1987,Pruschke_Anderson_1989,Pruschke_Hubbard_1993,Haule_Anderson_2001,Eckstein_Nonequilibrium_2010,hartle_decoherence_2013,Cohen_Greens_2014,Erpenbeck_Revealing_2021,Erpenbeck_Resolving_2021}.
At shorter timescales and smaller unit cell sizes, we cross-validate
results from these approximate schemes using the more computationally
expensive numerically exact inchworm Quantum Monte Carlo (iQMC) method
\citep{Cohen_Taming_2015}. The inchworm scheme takes advantage of
the causal structure of diagrammatic expansions to formulate resummed
Monte Carlo methods that bypass certain sign problems, including the
dynamical sign problem that usually limits nonequilibrium simulations
\citep{Antipov_Currents_2017,Chen_Inchworm_2017,Chen_Inchworm_2017_2,Boag_Inclusion_2018,ridley_numerically_2018,ridley_lead_2019,ridley_numerically_2019,krivenko_dynamics_2019,Cai_Inchworm_2020,Cai_Numerical_2020,kleinhenz_dynamic_2020,yang_inclusion-exclusion_2021,cai_fast_2022,kleinhenz_kondo_2022,kim_pseudoparticle_2022,li_interaction-expansion_2022,Erpenbeck_Quantum_2023}.

\paragraph{Diffusion.\textemdash{}}

In order to analyze localization we need to understand how the initial
state of the system, which is a spin density wave with wave number
$k$, evolves in time. We assume (and later test this assumption)
that in the hydrodynamic limit of $k\to0$ the system is well described
by the fractional diffusion equation \citep{metzler_random_2000},

\begin{equation}
\frac{\partial P_{n}\left(t\right)}{\partial t}=D_{\mu}\nabla^{\mu}P_{n}\left(t\right).\label{eq:diffusion_equation}
\end{equation}
Here $P_{n}\left(t\right)$ is the probability to have a spin up electron
on site $n$, $1<\mu<2$ is the dynamical exponent, $\nabla^{\mu}$
is the fractional Laplacian, and $D_{\mu}$ is the generalized diffusion
constant. It is more convenient to work in the Fourier domain of Eq.~(\ref{eq:diffusion_equation}),
$\frac{\partial P_{k}\left(t\right)}{\partial t}=-D_{\mu}\mid k\mid^{2\mu}P_{k}\left(t\right),$
where $P_{k}\left(t\right)=\stackrel[n=0]{l-1}{\sum}e^{-ikn}P_{n}\left(t\right)$
and $k\in\left\{ \left.\frac{\pi}{m}\right|m\in\left\{ 0,1,\ldots,l-1\right\} \right\} $.
The solution is then given by
\begin{equation}
P_{k}\left(t\right)=P_{k}\left(0\right)e^{-D_{\mu}\mid k\mid^{2\mu}t}.\label{eq:solution_of_diff_equ}
\end{equation}
This yields the following relation, which we use to extract the diffusion
exponent and diffusion constant from our simulations:
\begin{equation}
\ln\frac{P_{k}\left(t\right)}{P_{k}\left(0\right)}=-D_{\mu}\mid k\mid^{2\mu}t.\label{eq:log_of_solution}
\end{equation}
We consider initial states (\ref{eq:initial_state}), where only one
$k$-mode is excited.

\paragraph{Results.\textemdash{}}

\begin{figure}
\includegraphics{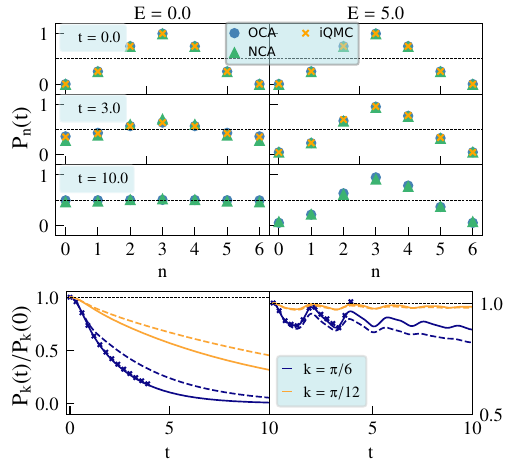}\caption{(upper panel) Time evolution of the spin up probability $P_{\mathrm{n}}\left(t\right)$
in equilibrium ($E=0)$ and with an electric field ($E=5$) for a
Coulomb interaction $U=2$, with circles/triangles/crosses denoting
NCA/OCA/converged iQMC results, respectively. The initial state is
a spin density wave with wavenumber $k=\pi/6$. (lower panel) Time
evolution of two Fourier components $k=\frac{\pi}{6},\frac{\pi}{12}$.
The dashed (solid) line denotes data obtained from NCA (OCA) calculations,
while symbols represent converged iQMC results.\label{fig:spin-populations}}
\end{figure}

The upper panels of Fig.~\ref{fig:spin-populations} show the spin
up population $P_{n}\left(t\right)$ at the 6 sites $n$ within a
unit cell of size $l=6$, at the initial time and several later times.
The initial state is a spin density wave characterized by $k=\frac{\pi}{6}$.
Different symbols denote the two approximations and the numerically
exact iQMC result, where available; this shows that the OCA is quantitatively
accurate at intermediate times. In the three left panels, the electric
field is turned off ($E=0$), and the spin density wave rapidly relaxes
to a uniform equilibrium state. In the three right panels, we set
the electric field to a high value ($E=5$), and the spin density
wave survives to rather long times, suggesting localization.

In the lower panels of Fig.~\ref{fig:spin-populations}, we plot
the Fourier transform of the spin up population, $P_{k}\left(t\right)$,
as a function of time for initial states with $k=\frac{\pi}{6}$ and
$\frac{\pi}{12}$. Dashed and solid lines correspond to the NCA and
OCA, respectively; and symbols to iQMC results, which are only evaluated
for the $k=\frac{\pi}{6}$ case. In the left panel, where $E=0$,
we observe a clear exponential decay for both values of $k$. On the
right, where $E=5$, we observe oscillatory dynamics that suggest
eventual decay for $k=\frac{\pi}{6}$. For $k=\frac{\pi}{12}$, however,
the population appears to freeze at a finite value, suggesting the
onset of localization. Notably, improving the approximation by going
from the NCA to the OCA enhances both trends. Furthermore, for $k=\frac{\pi}{6}$,
comparison between the OCA and the numerically exact iQMC results
shows that the OCA is accurate in this parameter regime.

\begin{figure}
\centering{}\includegraphics{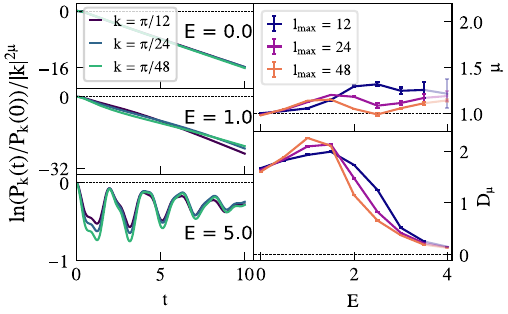}\caption{\emph{Left side}: each plot shows the decay of the populations for
three different initial states $k$ scaled by $|k|^{2\mu}$ after
the diffusion coefficient $\mu$ was estimated by fitting the data
to Eq.~(\ref{eq:log_of_solution}). The slope of the lines is the
diffusion constant $D_{\mu}$. \emph{Right side}: dynamical exponent
$\mu$ and the generalized diffusion constant $D_{\mu}$ for a system
with $U=2$ and different maximum unit cell sizes $l_{\mathrm{max}}$
used to extract $\mu,D_{\mu}$ (see main text for the definition).\label{fig:curve-collapse}}
\end{figure}

If the system obeys a generalized diffusion equation, Eq.~(\ref{eq:log_of_solution}),
curves like the ones in the bottom panels of Fig.~\ref{fig:spin-populations}
could be rescaled onto each other by plotting $\left|k\right|^{-2\mu}\ln\frac{P_{k}\left(t\right)}{P_{k}\left(0\right)}$
as a function of time $t$. The left panels of Fig.~\ref{fig:curve-collapse}
show how this works: for each field strength $E$ and at a particular
interaction strength $U=2$, we extract the dynamical exponent $\mu$
by fitting this single unknown parameter to minimize the minimal least
square distance between curves with different $k$. When the relaxation
is exponential, such that the physics is consistent with Eq.~(\ref{eq:log_of_solution}),
the curves collapse onto a single, straight line. The slope of this
line then uniquely determines the generalized diffusion constant $D_{\mu}$.
Curiously, the curve collapse still works even when the $k-$modes
do not exponentially decay in time (see bottom left panel of Fig.~\ref{fig:curve-collapse}
where $E=5$). In that case, however, the diffusion constant and exponent
are ill-defined and we do not present them. The borderline case is
plotted as slightly transparent.

\begin{figure}
\centering{}\includegraphics{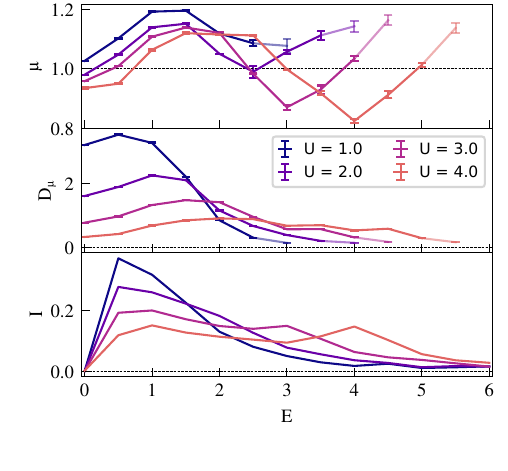}\caption{Dynamical exponent $\mu$ and generalized diffusion constant $D_{\mu}$
as functions of the electric field $E$, for various interaction strengths
$U$ and extracted from the data with minimal wave number, $k_{l_{\mathrm{max}}=48}$
. The lowest panel shows the current for a system initially prepared
in a Neél state at time $t=10$.\label{fig:Diffusion-coefficients}}
\end{figure}

Diffusion is a large wavelength phenomenon, therefore it is imperative
to examine our results in the $k\to0$ limit; yet numerically we can
only access finite values of $l$. In the right panels of Fig.~\ref{fig:curve-collapse}
we show the dynamical exponent $\mu$ (top) and the generalized diffusion
constant $D_{\mu}$ (bottom) as a function of the field $E$ for interaction
strength $U=2$. Both $\mu$ are $D_{\mu}$ are computed by collapsing
three sets of wave-vectors differing by the maximum unit cell size
used: $k_{l_{\mathrm{max}}=48}\in\left\{ \pi/48,\pi/24,\pi/12\right\} $,
$k_{l_{\mathrm{max}}=24}\in\left\{ \pi/24,\pi/12,\pi/6\right\} $
and $k_{l_{\mathrm{max}}=12}\in\left\{ \pi/12,\pi/6,\pi/3\right\} $\footnote{While we could use the largest unit cell size, $l=l_{\text{max}}$,
to compute all three values of $k$, this would be computationally
wasteful. Therefore we choose $l=\pi/k$ in our computations.}. The set $k_{l_{\mathrm{max}}=48}$ is therefore most characteristic
of the hydrodynamic limit $k\to0$. We see that convergence to $k\to0$
is obtained only at the small electric field limit, where transport
is clearly diffusive ($\mu\approx1$). However, a general trend emerges
at larger fields: at weak fields, transport slows down, becoming subdiffusive
$\left(\mu>1\right)$. This is accompanied by an increase in the diffusion
constant, which then drops at higher fields. Interestingly, transport
then becomes diffusive again when the electric field is of the order
of the interaction strength, $E\approx U$. For larger electric fields
transport becomes subdiffusive again and finally localizes at even
higher fields. In this large $E$ limit, Eq.~(\ref{eq:diffusion_equation})
is no longer satisfied, and the extraction of $\mu$ and $D_{\mu}$
loses its meaning (faded symbols in Figs.~\ref{fig:curve-collapse}
and \ref{fig:Diffusion-coefficients}). We note that convergence with
$l_{\mathrm{max}}$ is generally faster at higher $U$ \citep{supp}.

In the top two panels of Fig.~\ref{fig:Diffusion-coefficients} we
plot the dynamical exponent and the generalized diffusion constant,
as obtained from the from result sets analogous to $k_{l_{\mathrm{max}}=48}$
above, but for several values of the interaction strength $U$. The
general trend is as in the $U=2$ case, but for values of $U$ larger
than 2, transport briefly becomes superdiffusive (i.e. $\mu<1$) for
intermediate fields for which $E\approx U$. This is accompanied by
a drop in the generalized diffusion constant (middle panel). The super-diffusive
regime does not appear to vanish in the hydrodynamic limit, $k\to0$
limit (see \citep{supp}). We argue that this enhancement in transport
is related to meeting the resonance condition between the field and
the interaction energy of local single-charge excitations on one site.
At higher fields, localization eventually sets in, preventing the
interpretation as a diffusion equation (borderline cases are marked
by faded symbols). At $U=1$, this occurs immediately after the dip
at the resonance. At higher interaction strengths, it is preceded
by a second rise of $\mu$ towards subdiffusive behavior.

To show the emergence of localization, in the bottom panel of Fig.~\ref{fig:Diffusion-coefficients}
we show the current flowing through the system at the largest accessible
time $t=10$, after starting from a Neél state. This initial condition
is chosen for numerical convenience and is not crucial here (see \citep{supp}
for details). The current increases with field at small fields, then
reaches a plateau and begins to decrease, before finally vanishing
in the localized regime. However, at $U=3$ and $U=4$, an increase
is visible near the resonance condition, where superdiffusive exponents
are observed.

\paragraph{Discussion.\textemdash{}}

Using approximate and numerically exact methods, we studied the temporal
relaxation of density waves in the Hubbard model on an \emph{infinite}
Bethe lattice and in the presence of a constant electric field, $E$.
We found that for electric fields smaller than the interaction strength,
the magnitude of the waves relaxes exponentially, such that the density
satisfies a fractional diffusion equation with anomalous dynamical
exponent. We studied the dependence of the dynamical exponent on the
interaction strength $\left(U\right)$ and the strength of the electric
field $\left(E\right)$. For $U\leq2$, we find sub-diffusive behavior
for $E\lesssim U$, which crosses over to diffusion in the hydrodynamic
limit $\left(k\to0\right).$ For $E\gtrsim U$ there is no visible
decay of the density waves, accompanied with a vanishing current;
this is consistent with a transition from a diffusive metal to an
insulator. Our results suggest that the sub-diffusion, experimentally
observed in a two-dimensional system \citep{Guardado-Sanchez2020},
may result from the system being far from the hydrodynamic limit.

For $U>2$ the behavior is more peculiar. Here, we do not observe
a significant drift of the dynamical exponent in the hydrodynamic
limit. Moreover, it has a non-monotonous dependence on the electric
field $E$. For sufficiently small electric fields transport is diffusive
with a dynamical exponent $\mu=1$. Increasing the electric field
makes transport sub-diffusive as long as $E<U$. When the electric
field becomes comparable to the interaction strength, we observe a
noticeable acceleration of transport all the way to super-diffusion.
Further increasing the field, such that $E\gg U,$ leads to an apparent
localization. Working directly in the thermodynamic limit, our results
are consistent with Refs.~\citep{zisling2022:TransportStarkManybody,kloss2023:AbsenceLocalizationInteracting},
which suggest that Stark-MBL localization in a one-dimensional system
is possible only in the thermodynamic limit. Regarding future work,
we mainly focused on spin transport, but it would also be interesting
to contrast this to density transport. Another open question is the
nature of the super-diffusive transport for $E\approx U$, which we
associated with a resonance condition between the electric field and
the energy it takes to create or destroy a doublon/holon. It would
be especially interesting to realize this effect in cold atoms experiments.
\begin{acknowledgments}
This research was supported by the ISRAEL SCIENCE FOUNDATION (Grants
No. 2902/21, 1304/23 and No. 218/19) and by the PAZY foundation (Grant
No. 318/78). Until August 31, A.E. was funded by the Deutsche Forschungsgemeinschaft
(DFG, German Research Foundation) - 453644843. A.E. starting on September
1, and E.G., were supported by the U.S. Department of Energy, Office
of Science, Office of Advanced Scientific Computing Research and Office
of Basic Energy Sciences, Scientific Discovery through Advanced Computing
(SciDAC) program under Award Number DE-SC0022088. This research used
resources of the National Energy Research Scientific Computing Center,
a DOE Office of Science User Facility supported by the Office of Science
of the U.S. Department of Energy under Contract No. DE-AC02-05CH11231
using NERSC award BES-ERCAP0021805.
\end{acknowledgments}

\bibliographystyle{apsrev4-1}
\bibliography{manuscript}

\end{document}